# Astrometric Jitter as a Detection Diagnostic for Recoiling and Slingshot Supermassive Black Hole Candidates

Anavi Uppal[1] , Charlotte Ward[2] , Suvi Gezari[3,4] , Priyamvada Natarajan[1,5,6] , Nianyi Chen[7] , Patrick LaChance[7], and Tiziana Di Matteo[7]

[1] Department of Astronomy, Yale University, P.O. Box 208101, New Haven, CT 06520, USA
[2] Department of Astrophysical Sciences, Princeton University, Princeton, NJ 08544, USA
[3] Space Telescope Science Institute, 3700 San Martin Drive, Baltimore, MD 21218, USA
[4] Department of Physics and Astronomy, Johns Hopkins University, 3400 North Charles Street, Baltimore, MD 21218, USA
[5] Department of Physics, Yale University, P.O. Box 208121, New Haven, CT 06520, USA
[6] Black Hole Initiative, Harvard University, Cambridge, MA 02138, USA
[7] McWilliams Center for Cosmology, Department of Physics, Carnegie Mellon University, Pittsburgh, PA 15213, USA
Received 2024 May 16; revised 2024 August 27; accepted 2024 September 14; published 2024 November 8

## Abstract

Supermassive black holes (SMBHs) can be ejected from their galactic centers due to gravitational wave recoil or the slingshot mechanism following a galaxy merger. If an ejected SMBH retains its inner accretion disk, it may be visible as an off-nuclear active galactic nucleus (AGN). At present, only a handful of offset AGNs that are recoil or slingshot candidates have been found, and none have been robustly confirmed. Compiling a large sample of runaway SMBHs would enable us to constrain the mass and spin evolution of binary SMBHs and study feedback effects of displaced AGNs. We adapt the method of varstrometry—which was developed for Gaia observations to identify off-center, dual, and lensed AGNs—in order to quickly identify off-nuclear AGNs in optical survey data by looking for an excess of blue versus red astrometric jitter. We apply this to the Pan-STARRS1 $3\pi$ Survey and report on five new runaway SMBH candidates. We focus on ZTF18aajyzfv: a luminous quasar offset by $6.7 \pm 0.2$ kpc from an adjacent galaxy at $z = 0.224$, and conclude after Keck LRIS spectroscopy and comparison to ASTRID simulation analogs that it is likely a dual AGN. This selection method can be easily adapted to work with data from the soon-to-be commissioned Vera C. Rubin Telescope Legacy Survey of Space and Time (LSST). LSST will have a higher cadence and deeper magnitude limit than Pan-STARRS1, and should permit detection of many more runaway SMBH candidates.

*Unified Astronomy Thesaurus concepts:* Active galactic nuclei (16); Supermassive black holes (1663); Sky surveys (1464)

## 1. Introduction

Most local galaxies harbor a supermassive black hole (SMBH) in their centers (D. Richstone et al. 1998). When galaxies collide, their SMBHs will also merge to form a binary (M. C. Begelman et al. 1980), and their coalescence to form a more massive SMBH will be accompanied by the emission of gravitational waves. SMBHs are believed to grow both via gas accretion and mergers over cosmic time, while coevolving with their host galaxies (see review by P. Natarajan 2014) and references therein).

Special conditions—such as asymmetries in the mass ratio of the progenitor SMBHs, their spin magnitudes and directions, and their orbital parameters—can cause this gravitational-wave emission to be asymmetric, which can lead the newly merged SMBH to be kicked out of the center of the galaxy through the process of gravitational-wave recoil (M. Campanelli et al. 2007; L. Blecha & A. Loeb 2008; L. Blecha et al. 2011). If such an SMBH does not get completely ejected from its galaxy, it will remain bound and move through the galaxy and oscillate about the center of mass of the system, sinking back to the center of the gravitational potential over time due to dynamical friction. However, if such an SMBH continues to accrete matter, it may be visible as a bright, off-center active galactic nucleus (AGN) for approximately $10^4$–$10^7$ yr after recoil (L. Blecha & A. Loeb 2008; S. Komossa 2012).

Alternatively, if a third galaxy merges with a system that already contains a binary or dual AGN, a three-body encounter between the SMBHs can eject the lightest SMBH through the gravitational slingshot recoil mechanism and may accelerate the coalescence of the remaining binary pair (L. Hoffman & A. Loeb 2007; L. Blecha et al. 2019). These slingshot AGNs would appear similar to recoiling AGNs, with the distinction that slingshot AGN systems could contain multiple SMBHs and recoiling AGN systems are likely to comprise of only a single SMBH.

AGNs play a crucial role in galaxy evolution through the modulation of star formation due to energy injection via feedback processes (A. Cattaneo et al. 2009). If an AGN is completely ejected from its galaxy through gravitational-wave recoil or slingshot processes, the period of star formation may be extended and the stellar population would evolve differently (L. Blecha et al. 2011). Creating a large sample of runaway AGNs would enable us to study the effect of displaced AGN feedback on galaxies and to constrain the mass and spin evolution of binary SMBHs (L. Blecha et al. 2019a). Such candidates permit study of SMBH mergers prior to their direct detection as Laser Interferometer Space Antenna (LISA) sources in the future. We note that the runaway SMBHs are a subset of the predicted wandering populations of black holes that are also detected as offset SMBHs (A. Ricarte et al. 2021).

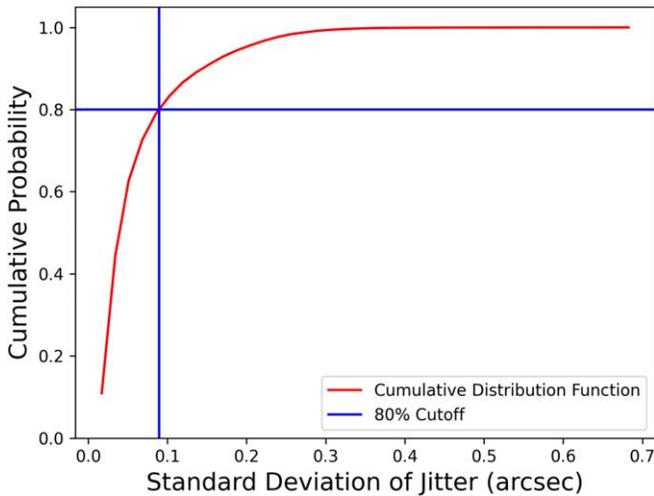

**Figure 1.** The $g$-band astrometric jitter for our control sample of AGNs from the Milliquas catalog. The 80% cutoff is at a jitter value of $0\rlap{.}{''}089$, and was adopted as our cutoff for selecting runaway AGNs in Pan-STARRS data.

However, despite the many important questions that runaway AGNs might address, only a handful of recoiling or slingshot AGN candidates have been found thus far, and none have been robustly confirmed (e.g., S. Komossa et al. 2008; M. Chiaberge et al. 2017; E. Kalfountzou et al. 2017; C. Ward et al. 2021; P. van Dokkum et al. 2023). Recoiling AGN candidates have previously been found through high-resolution imaging, forward modeling of variable AGNs in multiepoch imaging, and spectroscopy. The two main lines of evidence are the detection of AGN activity on kiloparsec-scale offset from galaxy nuclei and finding broad emission lines from the accretion disks that have a relative velocity greater than a few hundred $\mathrm{km\,s^{-1}}$ compared to left-behind narrow-line gas (M. Volonteri & P. Madau 2008). These identification methods are resource- and time-intensive, which contributes to why so few runaway AGN candidates have been identified thus far.

Distinguishing between spatially offset AGN and other phenomena can, however, be challenging, which likely also contributes to the low numbers of runaway AGN candidates. Velocity offsets between broad-line regions and narrow-line regions could be due to gas kinematics such as outflowing winds (J. T. Allen et al. 2015) rather than a runaway AGN. AGN-like variability can be mimicked by long-lived stellar transients, stellar outbursts, or supernovae. For example, SDSS J113323.97+550415.8—initially identified as a recoiling AGN candidate due to the presence of long-term optical variability—showed spectral features such as blueshifted absorption lines and Fe II and [Ca II] emission that are rare for an AGN, indicating that the source is most likely an erupting luminous blue variable star (M. Koss et al. 2014; C. Ward et al. 2021; M. Kokubo 2022). Without spatially resolved spectroscopy of an offset AGN and the putative host galaxy nucleus, it can be difficult to distinguish between truly offset AGNs and quasars which are blended with a foreground or background galaxy (C. Ward et al. 2021). AGN outflows, binary SMBH systems, and dual AGNs could also appear to be recoiling AGNs (e.g., J. Li et al. 2024). While a runaway AGN should carry a negligible stellar component, which should aid in distinguishing it from an inspiraling satellite galaxy with its own SMBH, it is possible for inspiraling SMBHs to also have mostly stripped stellar components. An estimated 8%–32% of SMBHs in the universe are in stripped nuclei (K. T. Voggel et al. 2019) and 0.5%–5% of AGNs are in kpc-scale binaries (Y. Shen et al. 2011a).

An existing method called "varstrometry" (merging the words variability and astrometry) uses detected excess astrometric noise in Gaia data (Gaia Collaboration et al. 2018) to flag unresolved off-center, dual, and lensed AGNs (Y. Liu 2015; Y. Shen et al. 2019; H.-C. Hwang et al. 2020; Y.-C. Chen et al. 2022b; E. Schwartzman et al. 2024). However, in order to take advantage of the large number of optical imaging surveys available today and in the future, a method that quickly identifies runaway AGN candidates in optical imaging survey data is desired. In particular, planned deep time-domain surveys such as the Vera C. Rubin Observatory's Legacy Survey of Space and Time (LSST) will enable the detection of offset AGNs at lower luminosities. LSST will survey the entire sky on the timescale of a week in *ugrizy* optical filters with an average $5\sigma$ depth for point sources of $r \sim 24.5$ in single exposures (Ž. Ivezić et al. 2019).

In this paper, we implement a novel technique that expands on varstrometry and only requires catalog-level products derived from optical imaging data to systematically identify runaway AGN candidates and can be easily scaled up for future surveys like LSST.

This paper is organized as follows. In Section 2, we introduce our selection method and apply it to data from the Pan-STARRS1 $3\pi$ Survey (K. C. Chambers et al. 2016), which obtained $\sim$10 epochs over a baseline of three years in *grizy* bands over 30,000 square degrees of sky. In Section 3, we detail Keck follow-up spectroscopy on our new candidate, i.e., ZTF18aajyzfv. In Section 4, we analyze the viability of ZTF18aajyzfv as a runaway AGN candidate by looking at the spectral features of the galaxy and AGN, and then performing point-spread function (PSF) subtraction. In Section 5, we look at theoretical counterparts of ZTF18aajyzfv in the ASTRID cosmological hydrodynamical simulation. In Section 6, we address the possible physical interpretations of ZTF18aajyzfv. In Section 7, we briefly discuss four other new runaway AGN candidates that we identified through our astrometric jitter selection method.

## 2. Identifying Runaway AGN Candidates

### 2.1. New Selection Method

To identify the presence of variable AGNs spatially offset from their host galaxies, we implemented a technique that adjusts the existing varstrometry method to be used with optical imaging survey data. Varstrometry can reveal the presence of offset AGNs from the astrometric jitter of a source, which refers to the variability in the observed position of a source (Y.-C. Chen et al. 2022b). This astrometric jitter is caused by competition between a constant-brightness galaxy nucleus (which could either be devoid of a central AGN or could host an obscured central AGN) and a variable-brightness offset AGN. AGNs are bluer and more variable than galactic centers or obscured central AGNs, and so we expect the $g$-band (bluer) detections of runaway AGNs to have more astrometric jitter than their $z$-band (redder) detections. We calculate the jitter by taking the standard deviation of the centroid separations in a given filter. For each filter, the separations were calculated by measuring the distances between the centroids measured in single-epoch detection images and the overall Pan-STARRS-reported mean centroid of the object, which was obtained from `raMean` and `decMean` in the `MeanObject` table.

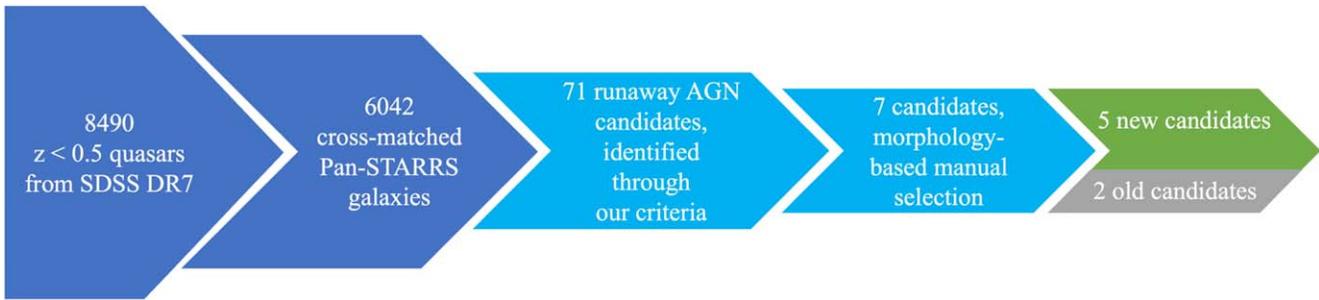

**Figure 2.** Flow chart detailing our filtering and selection process to identify runaway AGN candidates. We began with quasars identified in SDSS, found which ones corresponded to galaxies in Pan-STARRS, and then applied our astrometric jitter method to that sample. We then manually selected from the candidates that were provided by our selection method.

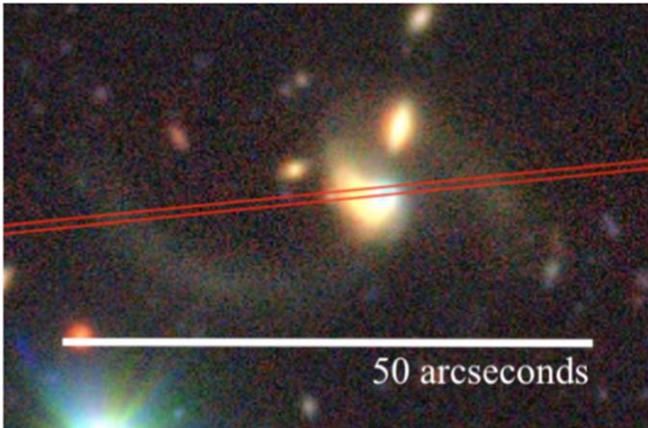

**Figure 3.** ZTF18aajyzfv's tidal arms, from Subaru Telescope Hyper Suprime-Cam Data Release 2. The tidal arms show that the system underwent a recent merger, which is necessary for classification as a runaway AGN. In red: Keck LRIS slit position from our 2023 June observations. It is angled 95°.45 east of north in order to capture both the AGN and galaxy in the same exposures.

In order to select cutoff values for the astrometric jitter, as well as ensure that we are detecting true astrometric jitter and not simply astrometric noise caused by detector or atmospheric effects, we calculate the $g$-band jitter of a control sample of AGNs chosen from the Million Quasars (Milliquas) Catalog (E. W. Flesch 2023). We expect the majority of this sample will be centrally located AGNs, with a small tail distribution of off-center AGNs that we want to be sensitive to through a jitter cutoff. This control AGN sample was limited to $z < 0.5$ because AGNs at higher redshifts generally do not have visible extended emission from a host galaxy. We then further selected only the quasars that were in the Pan-STARRS catalog and labeled as galaxies (using Pan-STARRS's recommended criteria of `iPSFMag - iKronMag` >0.05 mag to identify extended sources). Since noise from overly dim sources can cause false astrometric jitter, we also omitted AGNs that had $g$-band magnitudes dimmer than 18.94 mag. This cutoff was selected to match the cutoff we implemented while searching for new runaway AGNs in Section 2.2. Together, these cuts resulted in a control sample of 36,743 AGNs.

After performing three-sigma brightness clipping in each photometric filter in order to eliminate outlier detections, we find that 80% of our control AGN sample had a jitter of 0″089 or less, as shown in Figure 1. We use this as our cutoff $g$-band jitter value for selecting runaway AGN candidates. Although we did explore binning the control AGNs by $g$-band magnitude and having separate jitter cutoffs for each bin, the difference between the binned cutoffs was small and we decided that a universal cutoff was sufficient.

As a basic sanity check, we also add the criterion that candidates must have more blue ($g$-band) jitter than red ($z$-band) jitter. Finally, we implement the additional criterion that the $g$-band mean centroid position must be greater than 0″05 away from the $z$-band mean centroid position. We select this chromatic offset cutoff based on the initial results from implementing only the previous two criteria—the passing sample of AGNs contained a large number of still-merging galaxies, which are unlikely to be runaway AGN systems. Implementing a 0″05 chromatic offset cutoff helped mitigate this.

In order to assess if our selection criteria ensure reidentification of previously reported offset AGN candidates, we study nine previously identified offset AGNs that are chosen from Zwicky Transient Facility (ZTF) time-domain imaging by C. Ward et al. (2021). ZTF is a wide-field optical transient survey that images the northern sky in $g$-, $r$-, and $i$-bands with an average cadence of 2–3 days (E. C. Bellm et al. 2019; M. J. Graham et al. 2019). C. Ward et al. (2021) measured AGN–host spatial offsets by forward modeling time-resolved imaging from ZTF and deeper imaging from the DESI Legacy Imaging Survey (A. Dey et al. 2019) with The Tractor (D. Lang et al. 2016). The Tractor models the presence of a point source and a host galaxy in imaging, and the best-fit positions found in the model allowed C. Ward et al. (2021) to measure offsets between AGNs and their host galaxies.

We find that six of the nine runaway candidates in C. Ward et al. (2021) passed our 0″05 chromatic separation cutoff, providing us with a good balance between excluding merging galaxies and including runaway AGN candidates. Two of C. Ward et al.'s (2021) nine candidates passed our 0″089 $g$-band jitter cutoff. This means that the astrometric jitter selection method likely misses some runaway AGN candidates, but a stringent cutoff is necessary to omit central AGNs.

In summary, after performing three-sigma brightness clipping within each filter in order to remove outlier detections, our runaway AGN candidate selection criteria are:

1. $g$-band jitter > $z$-band jitter
2. $g$-band jitter >0″089
3. Average $g$-band centroid offset from average $z$-band centroid by >0″05

| Grism (Blue Side) | Grating (Red Side) | Dichroic | Seeing |
|---|---|---|---|
| 600/4000 | 400/8500 | 560 | ~0″78 |

We also considered using only the *g* versus *z* chromatic offset to select recoiling AGN candidates. However, this resulted in hundreds of potential candidates when tested on Pan-STARRS data. The vast majority of these did not look to be visually offset upon inspection. Evidently, *g*-band jitter must be included in the criteria to eliminate the majority of non-visually offset candidates.

### 2.2. Application of the Astrometric Jitter Method

We applied this selection method on Pan-STARRS data of a sample of AGNs selected from the Sloan Digital Sky Survey Data Release 7 (SDSS DR7) quasar catalog (Y. Shen et al. 2011b). Our filtering process is the same as for our control AGN sample, and is illustrated in Figure 2. We omitted AGNs that had a redshift greater than 0.5 because these generally did not have visible extended emission from a host galaxy and it is not possible to determine whether AGNs are spatially offset without the context of a visible host galaxy. This resulted in 8490 quasars. We then further selected only the quasars that were labeled as galaxies according to Pan-STARRS (using Pan-STARRS's recommended criteria of `iPSFMag - iKronMag` >0.05 mag to identify extended sources). We also omitted AGNs that were dimmer than C. Ward et al.'s (2021) nine previously identified runaway candidates, as noise from an overly dim source can cause false astrometric jitter, which eliminated AGNs with a mean *g*-band magnitude fainter than 18.94 mag. We then applied our astrometric jitter selection method to Pan-STARRS1 survey data on the remaining 6042 AGNs. This identified 71 possible recoiling AGN candidates.

We visually examined the astrometric jitter plots and Legacy color images of all 71 possible candidates to see if they looked similar to known runaway AGN candidates. In order to flag an object as a likely recoiling AGN candidate, we looked for a bright point source spatially offset from the center of the putative host galaxy.

Based on this visual analysis, we finally selected seven AGNs as runaway AGN candidates. Five of these had never been identified in the published literature before, while two of them—ZTF19aautrth and ZTF19aadggaf—were previously published as offset AGNs and possible runaway candidates by C. Ward et al. (2021). The characteristics of ZTF19aautrth and ZTF19aadggaf were used to create our astrometric jitter selection criteria, so our reidentification of them is unsurprising.

Of the five new runaway AGN candidates we identified, three were flagged as possible offset AGNs by C. Ward et al. (2021) based on ZTF imaging data, but were not presented in C. Ward et al.'s (2021) final sample of mergers/offset AGN because their offsets were not confirmed in the Legacy Survey image modeling stage of candidate selection (see Table 4, row 5, in C. Ward et al. 2021). These three candidates are 103913.81+094003.0 (ZTF19aavvwybq), 133636.65+420934.1 (ZTF18aajyzfv), and

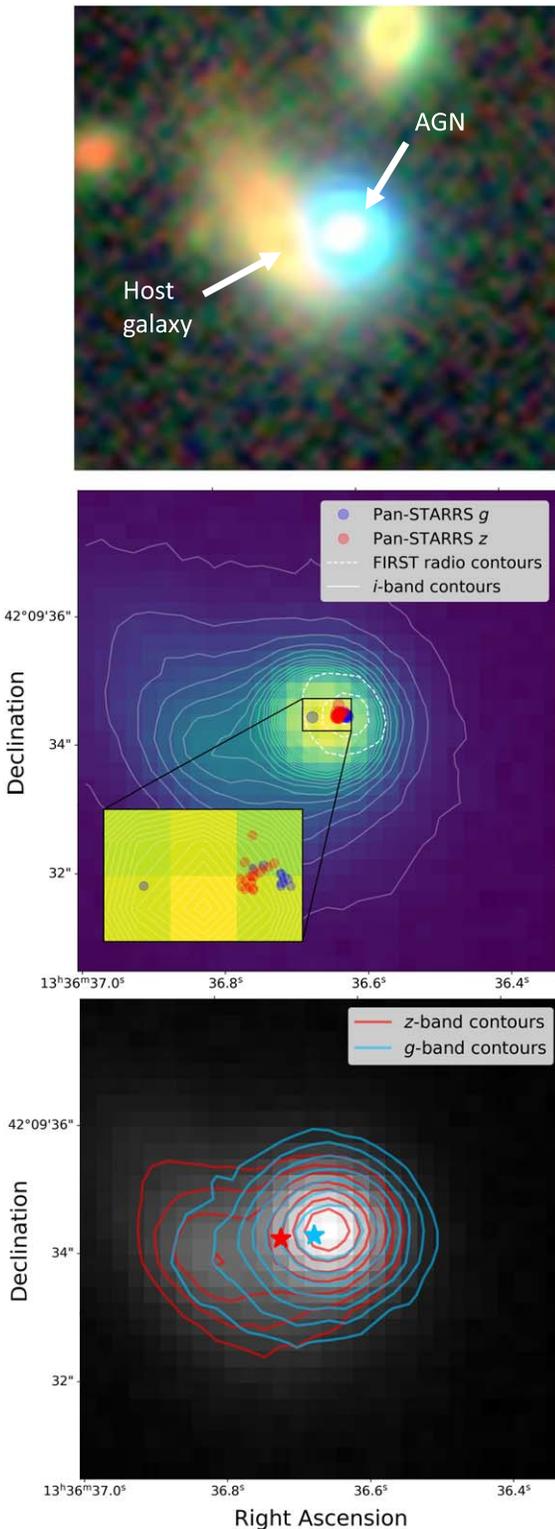

**Figure 4.** Top panel: Legacy Survey DR9 *grz* color image of ZTF18aajyzfv. Middle panel: The *g* and *z* detection centroid locations over time, overlaid on the Pan-STARRS1 *i*-band survey image, showing a greater standard deviation of the blue jitter compared to the red jitter. FIRST radio contours are overlaid with dotted lines, and are centered on the blue AGN. Bottom panel: Pan-STARRS1 *z* and *g* contours overlaid on the *i*-band image. *z* (red) and *g* (blue) "center of mass" centroids calculated from 2D image moments of Pan-STARRS stack images are marked with stars.

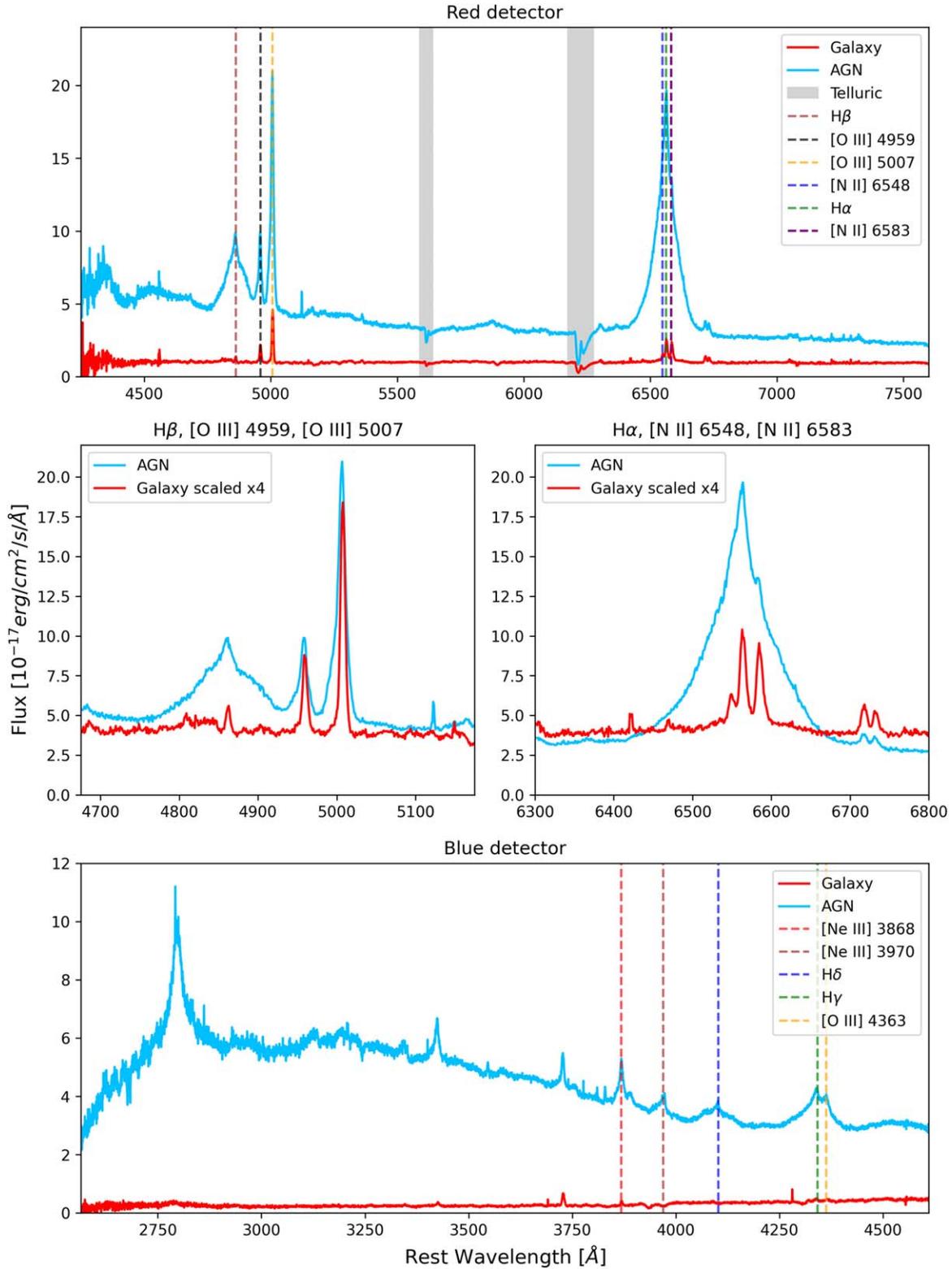

**Figure 5.** Keck LRIS spectra of ZTF18aajyzfv's AGN and galaxy. The AGN exhibits broad Balmer emission lines in addition to narrow lines, while the galaxy only exhibits narrow lines. The AGN spectrum has blueward bumps in its [O III] emission lines, indicating a possible outflow.

150829.92+451423.4 (ZTF18abkztag). C. Ward et al.'s (2021) unconfirmed candidates were not considered when creating our astrometric jitter selection criteria, and so our reidentification of them as runaway AGN candidates shows that our method successfully selects offset AGNs. Furthermore, as C. Ward et al.

(2021) identified these offset AGNs by using computationally intensive forward modeling of multiepoch imaging data, our reidentification of them using catalog-level data products shows that a simpler, less computationally intensive method can be used to identify the same candidates.

| R.A.<br>(deg) | Decl.<br>(deg) | AGN $z$ | Host $z$ | $g$-band<br>Jitter | Chromatic<br>Offset | $\log(M_{BH}/M_\odot)$ | Offset Galaxy<br>$r$ mag | Expected Host<br>$r$ mag |
|---|---|---|---|---|---|---|---|---|
| 204.15271 | 42.15950 | 0.224 | 0.224 | $0''10$ | $0''118 \pm 0''004$ | 8.38 | 16.5 | $18.8 \pm 0.2$ |

**Note.** The jitter and chromatic offset are calculated from Pan-STARRS catalog data, the redshifts are from our Keck LRIS spectroscopy, the offset galaxy magnitude is from our PSF subtraction, and the black hole mass is from H.-Y. Liu et al. (2019). The expected host magnitude refers to the expected magnitude of a galaxy that hosts the visible AGN in ZTF18aajyzfv, as derived in Section 4.4.

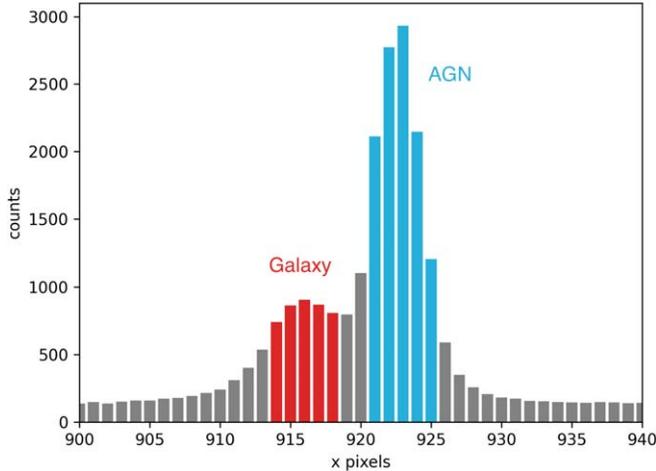

**Figure 6.** The 10-spectral-pixel average counts from a section of the red-detector Keck LRIS spectra, showing the overlap between the galaxy and AGN. We used 5 pixel extraction windows centered on the peak of the galaxy and AGN, indicated here in red and blue, respectively.

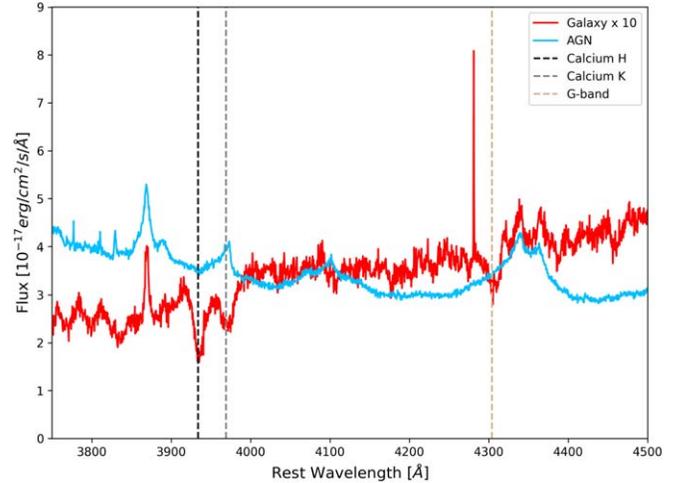

**Figure 7.** Stellar absorption lines in the spectrum of ZTF18aajyzfv's galaxy, shown in red, compared to the AGN's spectrum, shown in blue. The galaxy spectrum is scaled up by a factor of 10 to accentuate the absorption lines.

We identified ZTF18aajyzfv as a runaway AGN candidate worthy of further follow-up due to its clearly offset blue AGN, with no apparent extended emission centered on the AGN. This is the candidate that we will focus on for further detailed study in this paper. The other four new candidates are discussed in Section 8. ZTF18aajyzfv has a bright blue AGN offset from a redder galaxy by $6.7 \pm 0.2$ kpc, seen clearly in the top panel of Figure 4. The system's tidal arms, seen in Subaru imaging in Figure 3, reveal that this system has undergone a recent merger. As noted earlier, recoiling and slingshot AGNs form after merger events and can remain visible before tidal arms dissipate (L. Blecha et al. 2011).

ZTF18aajyzfv has more blue than red astrometric jitter, which is seen visually in the middle panel of Figure 4. Its AGN is also a radio source in the Very Large Array (VLA) FIRST Survey, shown in the same panel. Finally, the blue and red centroids of ZTF18aajyzfv are clearly offset from each other, as seen in the lower panel of Figure 4. The red and blue centroids are marked with stars in this panel, and are "center of mass" centroids calculated from 2D image moments. These centroids differ from the individual detection positions of red and blue centroids in ZTF18aajyzfv seen in the middle panel because we took the entire galaxy and AGN into account when calculating centroids, while Pan-STARRS used apertures centered on the AGN. Hence, our centroids are shifted more toward the galaxy when compared with Pan-STARRS's centroids.

Notably, our astrometric jitter method finds low-redshift runaway AGN candidates, while the original Gaia varstrometry method finds higher-redshift runaway AGN candidates. This is because the Gaia varstrometry method depends on the

Gaia `astrometric_excess_noise` parameter, and the extended structure of low-redshift galaxies can result in an artificially large `astrometric_excess_noise` that does not necessarily correspond to the jitter caused by competition between a stable-brightness galaxy nucleus and off-center AGN (H.-C. Hwang et al. 2020). Therefore, Y.-C. Chen et al. (2022b) excluded all $z < 0.5$ objects in order to omit all extended sources from their search for dual, off-nuclear, and lensed AGNs via varstrometry. In contrast, our method exclusively looks at $z < 0.5$ quasars in order to only consider extended sources. This means that these two different offset AGN selection methods provide a useful complement to each other in different search spaces over cosmic time.

## 3. Keck Follow-up Spectroscopy

In order to confirm that the AGN and offset galaxy of ZTF18aajyzfv are at the same redshift and to look for evidence of a velocity offset between the broad- and narrow-line emission regions of the AGN, we obtained spectra of the two components.

We observed ZTF18aajyzfv on June 21, 2023 UTC (NASA/Keck program 2023A_N076, PI: Ward) using the Low-Resolution Imaging Spectrometer (J. B. Oke et al. 1995; C. Rokosi et al. 2010) on the Keck I telescope at the W. M. Keck Observatory. LRIS splits light between a red detector and a blue detector. Together, they cover a wavelength range of 3200 to 10000 angstroms, have a pixel scale of 0.135 arcsec/pixel, and have a 6×7.8 arcmin field of view. We used a 1″ slit to obtain spectra of the ZTF18aajyzfv AGN and host galaxy together with the specifications in Table 1. For the red detector, we took three 1200-second exposures, and for the blue detector,

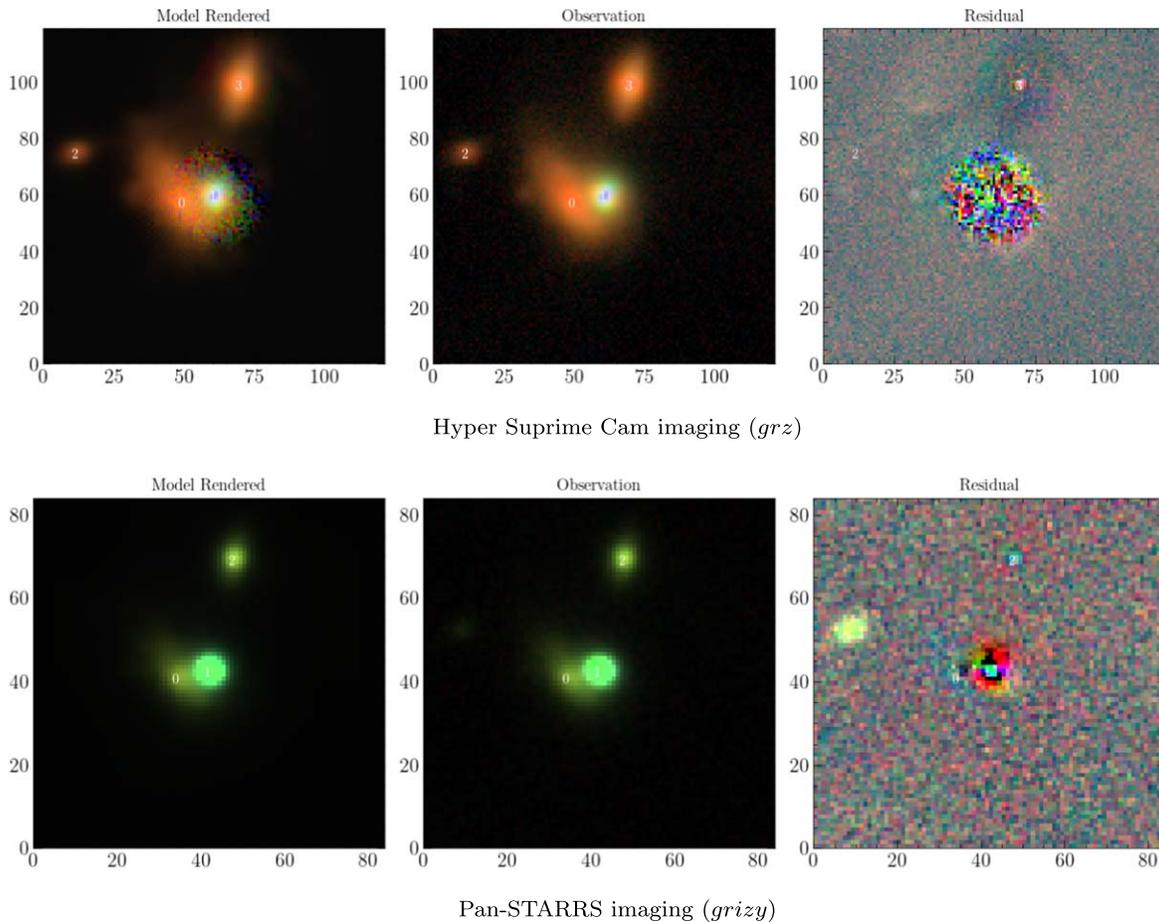

Hyper Suprime Cam imaging (*grz*)

Pan-STARRS imaging (*grizy*)

**Figure 8.** Top panels: `Scarlet` scene model from Hyper Suprime-Cam imaging. We show the model rendered to match the HSC imaging (left-hand panels), the coadded HSC *grz* image (center panels), and the residual (right-hand panels). The scene model consists of three extended sources (objects labeled 0, 2, and 3) and one point source coincident with the AGN (object 1). We note that the high white noise levels around the AGN in the HSC model and residuals arise due to high levels of noise in the PSF model, becoming the dominant source of pixel variance. Bottom panels: Same as above but for Pan-STARRS *grizy* imaging. Some artifacts due to an imperfect analytical PSF model are visible.

we took two 1800-second exposures, adding up to 3600 seconds of total integration time for each camera. Rather than observing at the parallactic angle, we angled the slit $95^\circ.45$ east of North to cover both the AGN and the host galaxy in the same exposure. The slit orientation is shown in Figure 3.

We manually separated and extracted the host and AGN traces, and then performed flux and wavelength calibrations using calculations obtained through the package PypeIt (J. X. Prochaska et al. 2020a, 2020b). These spectra are presented in Figure 5. As seen in the example extraction windows in Figure 6, the galaxy and AGN signals do slightly overlap, making it difficult to guarantee that the galaxy spectrum is not contaminated with AGN data, and vice versa. However, since the galaxy's spectrum does not seem to exhibit any broad-line emission bleedover from the AGN, we conclude that any contamination is likely minimal.

## 4. Analysis

### 4.1. Redshifts

The galaxy's stellar absorption features (shown in Figure 7) match the redshift of the AGN's emission lines. This confirms that the AGN and galaxy are at the same redshift. However, it does not rule out the possibility that the AGN actually resides in a faint second galaxy that is actively merging with the bright galaxy visible in the image.

We found the galaxy and AGN to be at $z = 0.224$, which is slightly different from SDSS's reported fiber spectrum redshift of 0.223. Upon inspection of the SDSS spectral fitting, we found that this is likely because of blueward wings in the AGN's [O III] emission lines (indicating possible outflows), which the SDSS fitting algorithm did not handle correctly.

### 4.2. Velocity Offsets

The spectrum of ZTF18aajyzfv's AGN reveals broad Balmer emission lines from high velocity gas close to the AGN, which are particularly visible in H$\alpha$ and H$\beta$. If ZTF18aajyzfv's AGN is a runaway SMBH and it has a velocity component in our viewing direction, we would expect to see a velocity offset between broad emission lines from gas carried away with the SMBH and the narrow emission lines left behind. This offset would be on the order of 100–1000 km s$^{-1}$.

Visually, it seems that the broad H$\beta$ line is offset to the right of the narrow H$\beta$ line. However, after continuum subtraction with the package `GELATO`[8] (Galaxy/AGN Emission Line Analysis TOol v2.4.1; R. E. Hviding et al. 2022), double

---

[8] https://github.com/TheSkyentist/GELATO

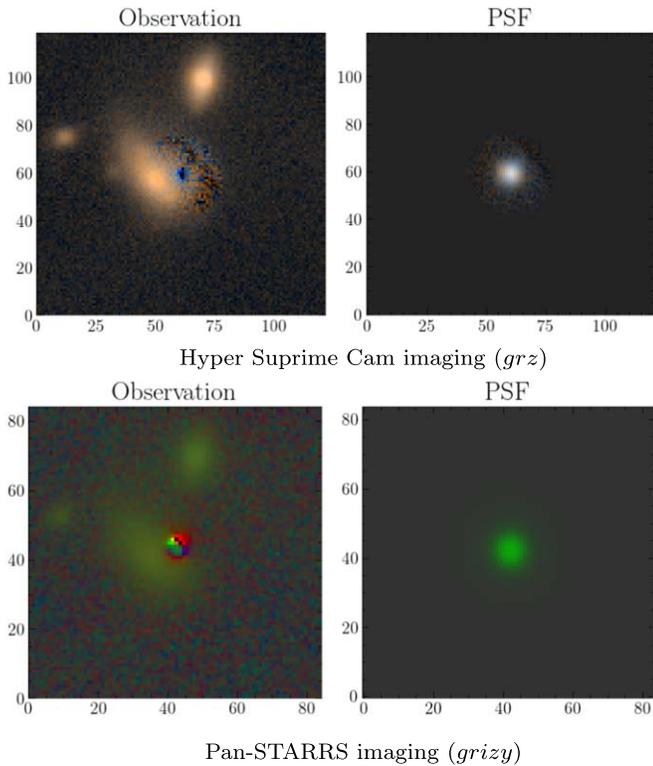

Observation · PSF

Hyper Suprime Cam imaging (*grz*)

Observation · PSF

Pan-STARRS imaging (*grizy*)

**Figure 9.** Top panels: *r*- and *z*-band HSC imaging of ZTF18aajyzfv with the best-fit point source model rendered to image space then subtracted from the data. The PSF model is shown on the right-hand side. Bottom panels: Same as above but for Pan-STARRS *grizy* imaging.

Gaussian modeling did not indicate a broad-line velocity offset greater than our velocity resolution limit of $62\,\mathrm{km\,s^{-1}}$. This does not completely eliminate the possibility of ZTF18aajyzfv being a recoiling or slingshot AGN—though unlikely, it is still possible that the velocity of the runaway SMBH is mostly orthogonal to our viewing direction, which would preclude us from detecting a significant radial velocity offset.

### 4.3. AGN PSF Subtraction

If ZTF18aajyzfv is not a runaway AGN, it could simply be a merging galaxy system. In this case, we would expect there to be a second galaxy in the system. In order to search for extended emission centered on AGN ZTF18aajyzfv indicating the presence of this second galaxy, we undertook modeling of the coadded *g*-, *r*-, and *z*-band imaging available from the Hyper Suprime-Cam Data Release 3 (HSC DR3; H. Aihara et al. 2022) (with a $0\farc17$ pixel scale) as well as the coadded *g*, *r*, *i*, *z*, and *y* Pan-STARRS imaging (with a $0\farc258$ pixel scale) using the `Scarlet` multiband scene modeling software[9] (P. Melchior et al. 2018). For the HSC models, we provided `Scarlet` with the PSF model images provided by HSC DR3, and modeled the system with four source components: a point source at the position of the variable AGN and a single extended source coincident with the spatially offset host galaxy, and two other extended sources to describe the background galaxies. We required that the extended galaxy models be monotonically decreasing—but not radially symmetric—and that they have the same morphology in each band (such that the SED does not vary in different regions of the galaxy). `Scarlet` was run until convergence criteria `e_rel` = $10^{-6}$ was met in order to fit the multiband SEDs, positions, and galaxy morphologies for the sources in the scene. For the Pan-STARRS imaging, we reconstructed the PSF image using the best-fit PSF parameters as described in E. A. Magnier et al. (2020) and published in the StackObjectAttributes table in the online Pan-STARRS DR1 catalog (H. A. Flewelling et al. 2020) and undertook the same `Scarlet` modeling procedure.

The best-fit model, corresponding observations, and residuals are shown in Figure 8. The HSC residuals do not show any evidence for extended emission from an additional host galaxy centered on the variable AGN, although the high level of background noise in the PSF model means that galaxy emission fainter than $g \sim 18.8$, $r \sim 19.0$ and $z \sim 18.9$ would be undetectable. The Pan-STARRS residuals show some evidence of additional extended emission around the AGN point source which has a similar color to the spatially offset galaxy nucleus. Assuming that up to 100% of residual emission could arise from an additional faint AGN-host galaxy instead of further extended emission from the spatially offset galaxy, we determine that the AGN may have extended galaxy emission of brightness less than or equal to $g \sim 18.5$, $r \sim 18.3$, $i \sim 18.0$, $y \sim 18.4$, and $z \sim 18.5$. In Figure 9, we show the HSC $r+z$ imaging after subtraction of the best-fit point source component only (we do not include the *g*-band image in the plot because an artifact close to the AGN center was masked) and the same for the Pan-STARRS *grizy* imaging. We use the `Scarlet` Pan-STARRS scene model to measure a spatial offset of $1\farc8 \pm 0\farc2$ between the best-fit subpixel position of the point source and the position of the brightest pixel in the galaxy morphology model. This corresponds to $6.7 \pm 0.2\,\mathrm{kpc}$ at $z = 0.224$.

### 4.4. Characteristics of a Possible Second Galaxy in ZTF18aajyzfv

In order to constrain the possible presence of secondary extended emission around the offset AGN, we predict the apparent magnitude of such a second galaxy by looking at the mass of the offset SMBH in ZTF18aajyzfv and then relating that mass to the galaxy magnitude.

H.-Y. Liu et al. (2019) estimated the virial mass of the SMBH in ZTF18aajyzfv by measuring the broad-line emission line width of H$\beta$ from an SDSS DR7 fiber spectrum centered on the AGN, which resulted in a value of $\log(M_{\mathrm{BH}}/M_\odot) = 8.38$. Then, using the following scaling relation from E. Tundo et al. (2007) between host galaxy magnitude and black hole mass:

$$\log(M_{\mathrm{BH}}/M_\odot) = (8.68 \pm 0.1) - \frac{(1.30 \pm 0.15)}{2.5}(M_r + 22)$$

we would expect a host galaxy with $M_r = -21.4 \pm 0.2$ mag for an SMBH of this mass, corresponding to $r = 18.8 \pm 0.2$ mag for a distance modulus of 45.25. The visible host galaxy in the top panel of Figure 4 has an *r*-band magnitude 16.5 mag, which is considerably brighter than the expected magnitude of $18.8 \pm 0.2$ mag.

Additionally, the upper limit on the residuals for a second galaxy underlying the offset AGN is $r = 18.3$ mag, as derived by our Pan-STARRS PSF modeling. Given this upper limit on the magnitude of a second host galaxy centered on the AGN, and the inconsistency between the visible galaxy's magnitude and the expected galaxy magnitude, we cannot rule out that the AGN is hosted in a second unseen galaxy.

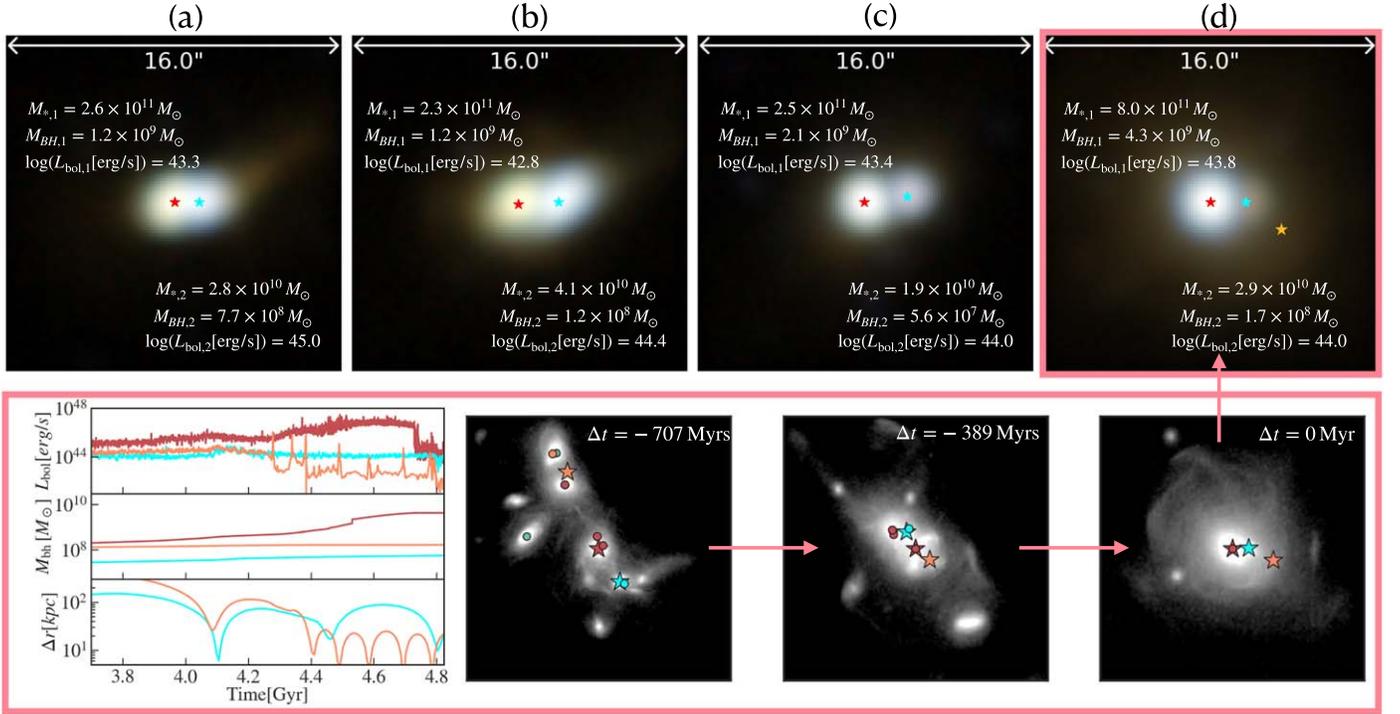

**Figure 10.** Analogs of ZTF18aajyzfv in the `ASTRID` cosmological simulation. Top row: Mock HSC Legacy Survey images for the four candidates from `ASTRID`. The red stars mark the position of the AGN in the central galaxies, and the cyan stars mark the position of the bright off-center AGN. System (d) is a triple MBH system with two AGN and one inactive, off-center MBH (labeled by the yellow star on the bottom right of the galaxies). Bottom row: Time evolution system (d). The left-hand panel shows the bolometric luminosities, masses, and separations of the three MBHs. The right-hand panels trace the evolution of the host galaxies and other MBHs in the surroundings. MBHs marked by the same colors will merge between the first and third frames.

## 5. Simulated Analogs from the ASTRID Simulation

To investigate the formation scenario of ZTF18aajyzfv, we search for theoretical counterparts of ZTF18aajyzfv in the `ASTRID` cosmological hydrodynamical simulation. `ASTRID` is a cosmological hydrodynamical simulation with $250h^{-1}$Mpc per side and $2 \times 5500^3$ initial tracer particles comprising dark matter and baryons (S. Bird et al. 2022; Y. Ni et al. 2022). The simulation includes full-physics, subgrid models for galaxy formation, SMBHs, and their associated supernova and AGN feedback, as well as inhomogeneous hydrogen and helium reionization. BHs are seeded in haloes with $M_{halo} > 5 \times 10^9 h^{-1} M_\odot$ and $M_* > 2 \times 10^6 h^{-1} M_\odot$, with seed masses stochastically drawn between $3 \times 10^4 h^{-1} M_\odot$ and $3 \times 10^5 h^{-1} M_\odot$, motivated by the direct collapse scenario proposed in G. Lodato & P. Natarajan (2007). The gas accretion rate onto the black hole is estimated via a Bondi-Hoyle-Lyttleton-like prescription (T. Di Matteo et al. 2005). The black hole radiates with a bolometric luminosity $L_{bol}$ proportional to the accretion rate $\dot{M}_*$, with a mass-to-energy conversion efficiency $\eta = 0.1$ in an accretion disk according to N. I. Shakura & R. A. Sunyaev (1973). In total, 5% of the radiated energy is coupled to the surrounding gas as the AGN feedback. Dynamics of the black holes are modeled with a subgrid dynamical friction model (M. Tremmel et al. 2015; N. Chen et al. 2022a), yielding well-defined black hole trajectories and velocities. Per this implementation, two black holes can merge if their separation is within two times the gravitational softening length $2\epsilon_g = 3$ ckpc $h^{-1}$ and their kinetic energy is dissipated by dynamical friction and they are gravitationally bound to the local potential.

We search for analogs of ZTF18aajyzfv in the $z \sim 1$ snapshot of `ASTRID`, with matched host galaxy mass (derived from the $r$-band magnitude), AGN luminosities, and projected separation. We found four close matches to ZTF18aajyzfv, and all of them are dual AGN systems, corroborating the theoretical interpretations provided in Section 6.3. We note, however, that in `ASTRID` we do not explicitly model three-body interactions between MBHs or gravitational-wave recoils after MBH coalescence, and we would only capture the dynamical interactions between multiple MBHs and their host galaxies on ~kpc scales (see e.g., Y. Ni et al. 2022, for details on the dynamics modeling). Therefore, cases of slingshot/recoil-induced analogs to ZTF18aajyzfv may exist but are missed in our search.

On the top row of Figure 10, we show the mock HSC Legacy Survey images for all four systems, with galaxy masses, black hole masses, and the AGN luminosities labeled for each AGN. In each image, the position of the AGN in the central galaxy is marked by a red star and a cyan star marks the position of the off-center luminous AGN. The mock images are made using a modified version of the synthetic observation pipeline described in P. LaChance et al. (2024). We assign stellar and AGN SEDs, place them in the image, and smooth them as described in that work. However, we adapted the pixel size, sensitivity, PSF, and other instrumentation to more closely match the Legacy Survey. All systems shown have a 2D projected separation between 5 and 7 kpc, and the 3D separations are (a) 5.0 kpc, (b) 14.9 kpc, (c) 8.6 kpc, and (d) 11.7 kpc. Among the four analogs, (a), (b), and (c) are dual AGN systems in the process of an ongoing galaxy merger. The central galaxy masses fall in the range $2-3 \times 10^{11} M_\odot$, with a central SMBH mass of $\sim 10^9 M_\odot$. The brighter AGN is embedded in a $2-4 \times 10^{10} M_\odot$ galaxy and has a bolometric luminosity of $\log(L_{bol}[{\rm erg\ s^{-1}}]) = 44 - 45$. All analogs have recently been through more than one galaxy merger, leaving a few less massive MBHs around the two SMBHs labeled in the mock images.

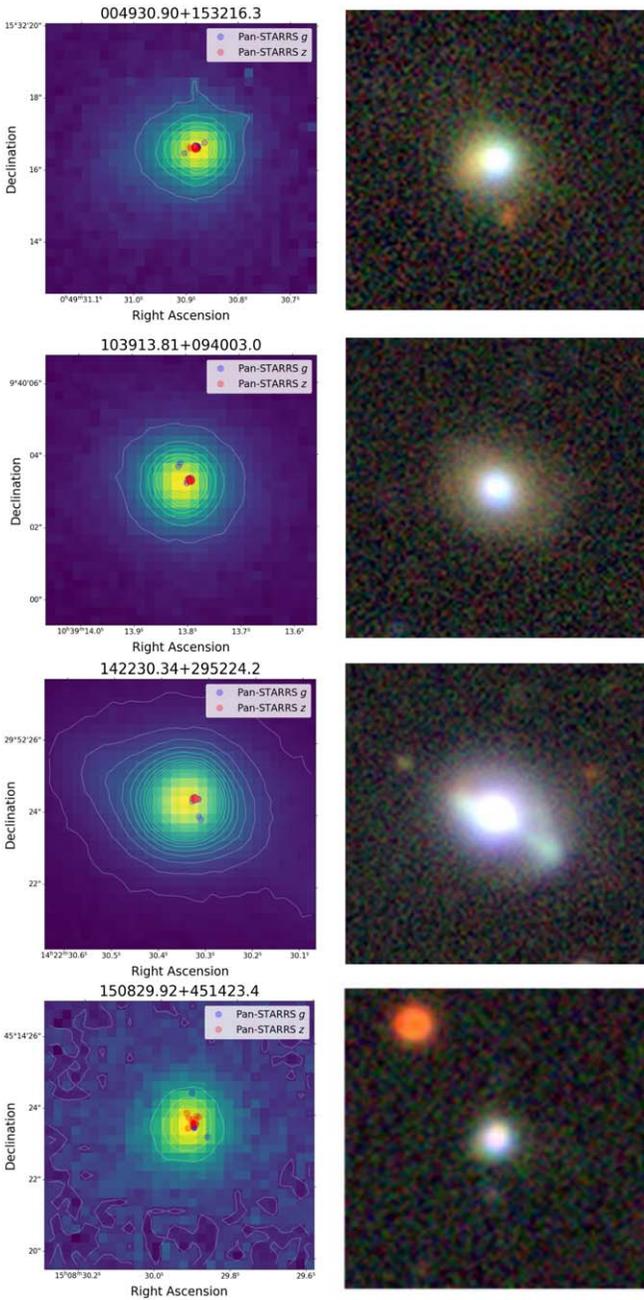

**Figure 11.** Four offset AGN candidates identified after visual inspection of candidates meeting the astrometric jitter criteria. Left-hand panels: $g$ and $z$ centroid locations over time, overlaid on the Pan-STARRS1 $i$-band survey image. Linear $i$-band contours are overlaid in white. Right-hand panels: Legacy Survey DR9 images.

System (d) is a triple MBH system with two AGN and one inactive, off-center MBH (labeled by the yellow star on the bottom right-hand side of the galaxies). The central galaxy and SMBH are a little brighter than the host galaxy in ZTF18aajyzfv, but this system resides in a complex environment and would most likely evolve into a gravitational slingshot with the three SMBHs involved if modeled further. We trace the evolution of this system over the past ∼700 Myr on the bottom row of Figure 10. This system results from a sequence of galaxy mergers involving more than five galaxies and AGN over the past ∼500 Myr. In the images showing the galaxy mergers, we marked the three MBHs in the mock images as stars with the corresponding colors, and the other MBHs with dots. The three MBHs marked with red dots all merged onto the central MBH (red star). The MBH marked with the cyan dot merged onto the off-center AGN (blue star), and the MBH marked by the orange dot merged with the third MBH (orange star). On the bottom left-hand side, we show the bolometric luminosities, masses, and separation of the three MBHs in case (d) over the past Gyr of evolution. The central MBH (red) has gone through a phase of rapid accretion following the three mergers it experienced, but then strong AGN outflow discontinues the phase of rapid accretion.

## 6. Possible Interpretations

We begin with three likely interpretations for the true nature of ZTF18aajyzfv: a recoiling AGN, a slingshot SMBH, or dual SMBHs. Deeper high-resolution imaging and/or IFU spectroscopy (e.g., from the Hubble Space Telescope or the James Webb Space Telescope) will be needed to definitively determine whether ZTF18aajyzfv is a runaway AGN or not. For example, CID-42 was thought to be a strong candidate for a recoiling or slingshot AGN until JWST spectroscopy revealed it to actually be a pair of merging galaxies (F. Civano et al. 2012; J. Li et al. 2024).

Here, we summarize current evidence for and against each possible interpretation.

### 6.1. Recoiling AGN

The narrow-line fluxes of the AGN's spectrum in ZTF18aajyzfv differ from the narrow-line fluxes in the offset galaxy's spectrum, indicating that we have resolved two separate narrow-line regions associated with the blue point source and the offset extended galaxy (Figure 5). The narrow-line fluxes observed from the offset extended galaxy indicate the presence of an additional active SMBH in its nucleus. The extended galaxy flux ratios of $0.82 \pm 0.05$ for [N II]6584/H$\alpha$ and $13.2 \pm 2.2$ for [O III]5007/H$\beta$ indicate that it is classified as a Seyfert 2 galaxy (J. A. Baldwin et al. 1981; L. J. Kewley et al. 2001), meaning that it hosts an AGN that has a broad-line region that is obscured by a torus (R. Antonucci 1993).

The existence of more than one SMBH in the system indicates that it is extremely unlikely for ZTF18aajyzfv to be a recoiling AGN system because a postrecoil system contains a single SMBH.

### 6.2. Slingshot SMBH

The presence of more than one SMBH still keeps alive the possibility that ZTF18aajyzfv is a slingshot SMBH system. Our search for analogs in the ASTRID simulation did find an analog gravitational slingshot system with host galaxy and AGN masses, AGN luminosities, and galaxy–AGN separations similar to ZTF18aajyzfv. However, due to the lack of a broad-line velocity offset, the inconsistency between the SMBH's mass and the offset galaxy's luminosity, and the possibility of underlying flux from a second galaxy, a slingshot scenario also seems unlikely.

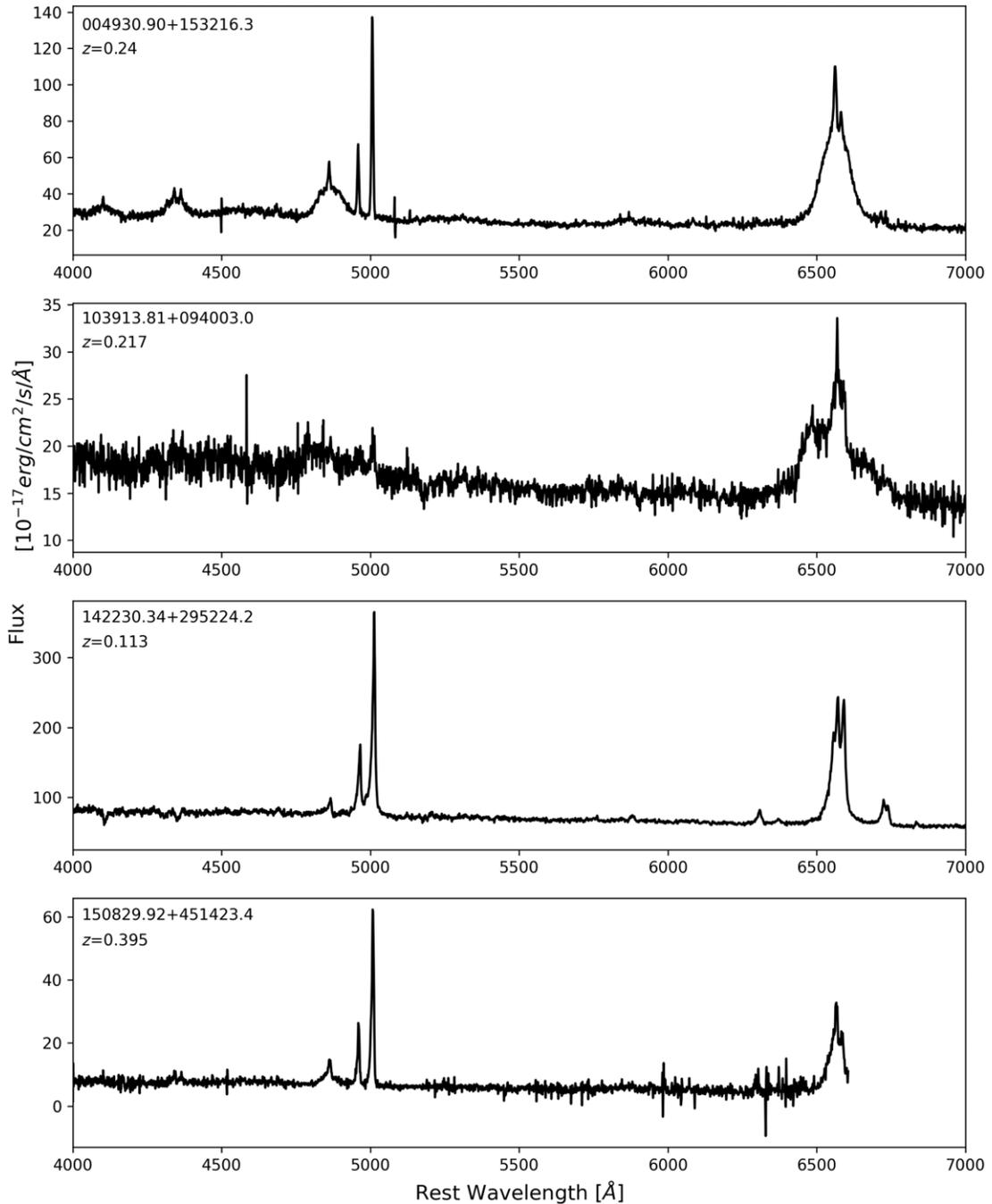

**Figure 12.** SDSS spectra of four additional offset AGN candidates. The first spectrum has broad emission lines that do not have a visible velocity offset and the second spectrum has broad emission lines that may have a blueward velocity offset indicative of a runaway AGN.

### 6.3. Dual AGNs

It is most likely that ZTF18aajyzfv is a late-stage merger with dual AGNs that are yet to coalesce or an early ongoing merger between two galaxies because our PSF subtraction could not rule out the existence of a second galaxy hosting the visible AGN. Moreover, ASTRID found three analogs to ZTF18aajyzfv that are dual AGN systems.

### 7. Four Additional New Runaway Candidates

We provide here a brief overview of the other four new runaway AGN candidates we identified using our method. Two of these were previously identified as possible offset AGNs in

ZTF imaging data but were not included among the final published candidates because they were not confirmed in an additional high-resolution imaging modeling step (103913.81 +094003.0 and 150829.92+451423.4; C. Ward et al. 2021). Astrometric jitter plots and Legacy color images are shown in Figure 11 and SDSS spectra are shown in Figure 12. Coordinates, redshifts, $g$-band jitter, and chromatic offset values are listed in Table 3.

1. 004930.90+153216.3 consists of a bright AGN clearly offset to the right of a potential host galaxy. It is not a radio source in Very Large Array Sky Survey (VLASS) data.

| Candidate | R.A. | Decl. | $z$ | Std. dev. of $g$-band Jitter | Chromatic Offset |
|---|---|---|---|---|---|
| 004930.90+153216.3 | 12.37875 | 15.53788 | $0.240 \pm 0.00002$ | $0\rlap{.}''09$ | $0\rlap{.}''123 \pm 0\rlap{.}''008$ |
| 103913.81+094003.0 | 159.80753 | 9.66755 | $0.217 \pm 0.00011$ | $0\rlap{.}''16$ | $0\rlap{.}''082 \pm 0\rlap{.}''007$ |
| 142230.34+295224.2 | 215.62644 | 29.87338 | $0.113 \pm 0.00002$ | $0\rlap{.}''18$ | $0\rlap{.}''055 \pm 0\rlap{.}''003$ |
| 150829.92+451423.4 | 227.12471 | 45.23984 | $0.395 \pm 0.00005$ | $0\rlap{.}''27$ | $0\rlap{.}''143 \pm 0\rlap{.}''014$ |

**Note.** Redshifts are from SDSS fiber spectra.

2. 103913.81+094003.0 (ZTF19aavwybq) consists of a bright AGN with a sufficiently small spatial offset from its putative host galaxy that it is not visible by eye, but it is statistically offset based on astrometric jitter and chromatic offset. Its SDSS spectrum in the second panel of Figure 12 shows a peak visible at approximately 6450 Å, blueward of its broad Hα emission line. This could be due to several phenomena, one of which is the velocity offset of broad-line gas in a runaway AGN's accretion disk. However, the shape of the broad-line profile is also consistent with double-peaked emission from the Keplerian motion of gas in an accretion disk, where Doppler boosting results in a brighter blue peak compared to the red peak (see M. Eracleous et al. 2009, for a review). Double-peaked accretion disk profiles are observed in 2%–10% of broad-line AGN, but were identified in the majority of the offset AGN identified in ZTF by C. Ward et al. (2021), perhaps suggesting a connection between recent merger activity and the occurrence of disk-dominated broad-line emission. In summary, due to multiple confirmations of a statistical spatial offset in this work and in C. Ward et al. (2021), and a possible broad-line velocity offset, 103913.81+094003.0 is a prime candidate for further follow-up with integral-field spectroscopy and high-resolution imaging in order to confirm or rule out this object as a runaway SMBH system.

3. 142230.34+295224.2 has a visibly offset AGN that is a radio source in VLASS. There are green extended sources on either side of the AGN, but neither are radio-bright.

4. 150829.92+451423.4 has a visibly offset AGN and is not a radio source in VLASS. In unpublished data, C. Ward et al. (2021) visually identified 103913.81+094003.0 as a possible coincidence between a foreground/background AGN and a background/foreground galaxy. Follow-up spectroscopy would be needed to confirm whether or not the AGN and galaxy are at the same redshift.

## 8. Summary and Conclusions

Creating a large sample of runaway AGN candidates will allow us to study the effect of displaced AGN and their feedback during galaxy mergers and—in the case of confirmed gravitational-wave recoil—also allow us to constrain the mass and spin evolution of binary SMBHs. We describe and implement a new method to identify recoiling and gravitational slingshot AGN candidates by looking at red versus blue astrometric jitter in optical survey data. We have derived the following selection criteria for Pan-STARRS imaging data: the standard deviation of the jitter of a candidate's $g$-band coordinates must be $>0\rlap{.}''089$, it must have more $g$-band jitter than $z$-band jitter, and the $g$ and $z$ stacked centroids of the coordinates must be separated by more than $0\rlap{.}''05$. This method is an expansion on the existing varstrometry method for identifying offset, dual, and lensed AGNs in Gaia data. While Gaia varstrometry finds candidates at $z > 0.5$, the optical imaging astrometric jitter method described in this paper finds candidates at $z < 0.5$, providing a useful complement.

Our method has been successful at identifying low-redshift offset AGNs in Pan-STARRS optical data. Through this method, we identified five new runaway AGN candidates. Three of these were previously identified as possible offset AGNs via modeling of multiepoch ZTF imaging data in C. Ward et al. (2021), lending credibility to our alternative method's selection capabilities. We performed Keck LRIS spectroscopy and detailed follow-up analysis on ZTF18aajyzfv, a luminous quasar at $z = 0.224$ offset from an extended galaxy by $6.7 \pm 0.2$ kpc. Through this analysis and a search for system analogs in the `ASTRID` simulation, we conclude that ZTF18aajyzfv is likely a dual AGN system and not a runaway AGN, though high-resolution imaging from JWST or Hubble will be necessary to completely rule out the gravitational slingshot AGN scenario.

This new method is also less time- and resource-intensive than the currently deployed methods for identifying runaway AGNs in optical imaging data because it requires only catalog-level data astrometric data products from multiepoch imaging. In the future, using this astrometric jitter method on deep optical imaging data from Rubin Observatory's LSST should permit the assembly of a sample of many recoiling and slingshot AGN candidates.

## Acknowledgments

We thank the anonymous referee for thoughtful comments which helped improve this paper.

This work was supported by a NASA Keck PI Data Award, administered by the NASA Exoplanet Science Institute. Data presented herein were obtained at the W. M. Keck Observatory from telescope time allocated to the National Aeronautics and Space Administration through the agency's scientific partnership with the California Institute of Technology and the University of California. The Observatory was made possible by the generous financial support of the W. M. Keck Foundation. This research has made use of the Keck Observatory Archive (KOA), which is operated by the W. M. Keck Observatory and the NASA Exoplanet Science Institute (NExScI), under contract with the National Aeronautics and Space Administration.

The Pan-STARRS1 Surveys (PS1) and the PS1 public science archive have been made possible through contributions by the Institute for Astronomy, the University of Hawaii, the Pan-STARRS Project Office, the Max-Planck Society and its participating institutes, the Max Planck Institute for Astronomy, Heidelberg and the Max Planck Institute for Extraterrestrial Physics, Garching, The Johns Hopkins University, Durham University, the University of Edinburgh, the Queen's University Belfast, the Harvard-Smithsonian Center for Astrophysics, the Las Cumbres Observatory Global Telescope Network Incorporated, the National Central University of Taiwan, the Space Telescope Science Institute, the National Aeronautics and Space Administration under grant No. NNX08AR22G issued through the Planetary Science Division of the NASA Science Mission Directorate, the National Science Foundation grant No. AST-1238877, the University of Maryland, Eotvos Lorand University (ELTE), the Los Alamos National Laboratory, and the Gordon and Betty Moore Foundation.

A.U. acknowledges support from the Research Experiences for Undergraduates program at the Nantucket Maria Mitchell Association, funded by NSF REU grant AST-2149985.

P.N. acknowledges support from the Gordon and Betty Moore Foundation and the John Templeton Foundation that fund the Black Hole Initiative (BHI) at Harvard University where she serves as one of the PIs.

*Facility:* Keck:I (LRIS)

*Software:* astropy (Astropy Collaboration et al. 2013, 2018), GELATO (R. E. Hviding et al. 2022), PypeIt (J. X. Prochaska et al. 2020a, 2020b), Scarlet (P. Melchior et al. 2018).

## ORCID iDs

Anavi Uppal  https://orcid.org/0000-0001-7658-9710
Charlotte Ward  https://orcid.org/0000-0002-4557-6682
Suvi Gezari  https://orcid.org/0000-0003-3703-5154
Priyamvada Natarajan  https://orcid.org/0000-0002-5554-8896
Nianyi Chen  https://orcid.org/0000-0001-6627-2533
Tiziana Di Matteo  https://orcid.org/0000-0002-6462-5734